\documentclass[twocolumn]{aastex631}

\shorttitle{Fluorine Abundances in Local Stellar Populations}
\shortauthors{Brady et al.}
\graphicspath{{./}{figures/}}
\DeclareUnicodeCharacter{2212}{-}

\begin{document}

\title{Fluorine Abundances in Local Stellar Populations}

\author[0000-0003-3271-9434]{K. E. Brady}
\affiliation{Indiana University Bloomington, Department of Astronomy, 727 East 3rd St., Bloomington, IN 47405, USA}

\author[0000-0002-3007-206X]{C. A. Pilachowski}
\affiliation{Indiana University Bloomington, Department of Astronomy, 727 East 3rd St., Bloomington, IN 47405, USA}

\author[0000-0001-7366-7699]{V. Grisoni}
\affiliation{INAF, Osservatorio Astronomico di Trieste, via G.B. Tiepolo 11, I-34131, Trieste, Italy}

\author[0000-0002-0475-3662]{Z. G. Maas}
\affiliation{Indiana University Bloomington, Department of Astronomy, 727 East 3rd St., Bloomington, IN 47405, USA}

\author[0000-0001-8383-5115]{K. A. Nault}
\affiliation{University of Iowa, Department of Physics and Astronomy, 203 Van Allen Hall, Iowa City, IA 52242, USA}



\begin{abstract}
We present the first fluorine measurements in 12 normal giants belonging to the Galactic thin and thick disks using spectra obtained with the Phoenix infrared spectrometer on the 2.1m telescope at Kitt Peak. Abundances are determined from the (1$-$0) R9 2.3358 $\mu$m feature of the molecule HF. Additionally, sodium abundances are derived in 25 giants in the thin disk, thick disk, and halo using the Na I line at 2.3379 $\mu$m. We report fluorine abundances for thin and thick disk stars in the metallicity range $-$0.7 $<$ [Fe/H] $<$ 0. We add two abundance measurements for stars with [Fe/H] $<$ 0.5 dex which are at a critical metallicity range to constrain models. We find a larger dispersion in fluorine abundances than sodium abundances despite both species having similar overall uncertainties due to atmospheric parameters, suggesting this dispersion is real and not observational. The dispersion is slightly larger in the thick disk than the thin. The thin and thick disk average [F/Fe] for our sample of stars combined with the literature differ by 0.03 dex. The observations are compared to available chemical evolution models.
\end{abstract}


\section{Introduction} \label{sec:intro}

The abundances of light elements provide insight into the chemical evolution of the Milky Way and its stellar populations (see \citealt{matt21} for a review). The cosmic origin of light elements with even atomic numbers (e.g., O, Mg, Si) is generally well understood, but the origin of many odd-$Z$ elements (e.g., F, P, Cl) remains under debate \citep{NKT13}. Understanding the formation of odd elements provides an opportunity to constrain nucleosynthesis processes as, for example, $^{19}$F (the lone stable isotope of fluorine) is integral in proton-capture reactions that include the CNO and Ne$-$Na cycles.

The light, odd-$Z$ element fluorine is thought to be produced in multiple stellar sites such as asymptotic giant branch (AGB) stars \citep{JOR}, rapidly rotating massive stars \citep{Pran18}, Wolf$-$Rayet stars \citep{MA00}, the $\nu$ process in core-collapse supernovae \citep{WOOS1}, and novae \citep{Olive19}. In the context of Galactic chemical evolution, several nucleosynthetic channels may contribute to the Galactic fluorine abundance, with different processes dominating certain stellar populations (such as thin-disk, thick-disk, and bulge stars) and/or at certain metallicities. For a comprehensive review of the proposed nucleosynthetic sources of fluorine, see Section 5.1 in \citet{Ryde20}.

Determining the abundance of fluorine in stars remains difficult due to many observational challenges. Due to the lack of strong, measurable atomic fluorine lines, the only indicator of fluorine abundance that can be measured in cool atmospheres are spectral lines from the hydrogen fluoride (HF) molecule. Vibration-rotation HF lines exist in the $K$ band, and pure rotational HF lines in the $N$ band. Telluric lines in the near-infrared (near-IR) can complicate the derivation of abundances, and the HF molecule has a relatively low dissociation energy, limiting studies to stars below $\sim$4700 K.

While early work on the abundance of fluorine in stars focused primarily on AGB stars
\citep[e.g.,][]{JOR,CUN,Abia09,Abia10} and carbon-enhanced metal-poor stars \citep{S07, LUC}, recent investigations have turned to the determination of the fluorine abundance in main-sequence and pre-AGB giant stars to identify the source of fluorine in the Milky Way. These studies, such as \citet{Jon14,Jon17}, \citet{PP}, \citet{Guer19}, and \citet{Ryde20}, use high-resolution spectrometers in the near-IR to measure HF at 2.3 $\mu$m. For the first time in abundance derivation studies, \citet{Jon14b} explored a pure rotational HF line at 12.2 $\mu$m, finding that the derived fluorine abundance agreed with the 2.3 $\mu$m vibrational-rotational HF line. Many of these studies have contrasting conclusions regarding the chemical evolution of fluorine in the Milky Way. Most recently, \citet{Ryde20} indicated a need for multiple cosmic sources of fluorine over time.

The solar abundance of fluorine was originally determined to be log $\epsilon$(F) = 4.56 $\pm$ 0.30 by \citet{HN69} based on observations of sunspots.  Since then, new measurements of the oscillator strengths and excitation potentials of the IR (1$-$0) HF features have become available in the HITRAN database \citep{Roth13}, and \citet{Mai14} have reanalyzed the solar abundance of fluorine. With the new, experimental molecular data, \citet{Mai14} found a solar fluorine abundance of log $\epsilon$(F) = 4.40 $\pm$ 0.25 from a careful analysis of eight HF features in sunspot umbral spectra from the \citet{WL01} atlas. This value is in good agreement with the meteoritic abundance at log $\epsilon$(F) = 4.42 $\pm$ 0.06 \citep{Lodders09}.

Just as new observational results became available, so too have new chemical-enrichment models that provide a theoretical framework for interpreting the abundance of fluorine in the Galactic disk. Authors include different nucleosynthesis prescriptions for fluorine to predict the abundance of fluorine as a function of [Fe/H]\footnote{We use the standard spectroscopic notation where [$A/B$] $\equiv$ log(N$_{\rm A}$/N$_{\rm B}$)$_{\rm star}$ $-$ log(N$_{\rm A}$/N$_{\rm B}$)$_{\sun}$ and log$\epsilon$(A) $\equiv$ log(N$_{\rm A}$/N$_{\rm H}$) + 12.0 for elements A and B.} and [O/H]. Recent studies include \citet{Kob+a,Kob+20}, \citet{Spit18}, \citet{Pran18}, \citet{Olive19}, \citet{gris20}, and \citet{Womack23}.

In this work, we report on new determinations of the abundance of fluorine in giants belonging to the Galactic thin and thick disks, and compare the dependence of [F/Fe] on [Fe/H] and [F/O] on [O/H] both with the thin and thick disks and with theoretical models. In Section \S\ref{obs} we describe the target selection, observations, and data reduction. Section \S\ref{analysis} details the spectral analysis. We compare our results to the literature in Section \S\ref{comparison}. In Sections \S\ref{disks} and \S\ref{evolution} we discuss fluorine enrichment in the thin and thick disks and the chemical evolution of fluorine. Finally, in Section \S\ref{summary} we summarize our results.

\begin{deluxetable*}{lrllrlrr}
\tabletypesize{\scriptsize}
\tablecaption{Stars Observed with Phoenix and the Kitt Peak 2.1 m Telescope\label{table_log}}
\tablewidth{0pt}
\tablehead{
   & & & \colhead{Spec.} & &  \colhead{UT} & \colhead{Exp.} & \colhead{S/N}  \\
   \colhead{Star} & \colhead{HIP} &  \colhead{Alt.} & \colhead{Type} &
   \colhead{K Mag.} & \colhead{Date} & \colhead{Time (s)} & \colhead{Ratio} 
}
\startdata
\multicolumn{8}{c}{\textbf{Thin Disk}} \\
HD 6555 & 5213 & BD +22 174 & K0III & 5.410 &  28 Dec. 2013 & 4 x 600  & 160  \\
HD 10057 & 7632  & BD +1 293  & K2III & 3.059 & 30 Dec. 2013  & 4 x 300 & 300 \\
HD 29065 & 21297 & HR 1452    & K3III & 1.820 & 2 Dec. 2012   & 4 x 60  & 120  \\
HD 49520 & 32844 & 58 Aur     & K3III & 2.024 & 28 Dec. 2013  & 4 x 90  & 200 \\
HD 106760 & 59856 & HR 4668 & K0.5III & 2.110 & 29 Dec. 2013 & 4 x 90 & 300  \\
HD 211683 & 110141 & BD +09 5019 & K2 & 4.612 & 31 Dec. 2013 & 4 x 400 & 205 \\
HD 212074 & 110382 & BD +14 4782 & K1IV & 5.013 & 1 Jan. 2014 & 4 x 600 & 200 \\
HD 216174 & 112731 & HR 8688 & K1III & 2.625 & 30 Dec. 2013 & 4 x 150 & 195 \\
HD 233517 & \nodata & BD +53 1239 & K2 & 6.637 & 30 Nov. 2012 & 4 x 900 & 160 \\
\multicolumn{8}{c}{\textbf{Thick Disk}} \\
HD 249   & 607 & BD +25 5073 & K1IV & 4.879 & 28 Dec. 2013 & 4 x 600  & 190  \\
HD 3546 & 3031 & $\epsilon$ And & G7III  & 2.070 & 28 Dec. 2013 & 4 x 120 & 260 \\
HD 9138  & 7007  & $\mu$ Psc & K3III & 1.660 & 30 Nov. 2012  & 4 x 30  & 210 \\
HD 15596 & 11698 & 27 Ari & G8III & 3.861 & 31 Dec. 2013 & 4 x 400 & 300  \\
HD 37171 & 26386 & HR 1908    & K5III & 2.120 & 28 Dec. 2013  & 4 x 90  & 145 \\
HD 44030 & 30142 & BD +25 1223 & K4III & 4.139 & 28 Dec. 2013  & 4 x 300 & 300 \\
HD 50778 & 33160 & tet CMa    & K4III & 0.660 & 30 Nov. 2012  & 4 x 20  & 130 \\
HD 71597 & 41538 & BD +0 2305 & K1IV & 4.503 & 1 Dec. 2012   & 4 x 500 & 190 \\
HD 77729 & 44592 & BD +26 1895 & K4III & 4.170 & 30 Dec. 2013 & 4 x 400 & 160 \\
HD 96436 & 54336 & 65 Leo &  K0III  &  3.329  & 29 Dec. 2013 & 4 x 300 & 240 \\
HD 103813 & 58306 & BD +27 2073 & K0III & 5.057 & 28 Dec. 2013 & 4 x 600 & 125 \\
HD 105740 & 59334 & BD +17 2445 & G9III & 5.847 & 29 Dec. 2013 & 4 x 600 & 165 \\
HD 107328 & 60172 & c Vir & K0.5III & 2.200 & 29 Dec. 2013 & 4 x 90 & 180 \\
HD 121146 & 67589 & HR 5227 & K2IV & 3.655 & 17 Feb. 2014 & 4 x 300 & 200 \\
HD 221345 & 116076 & 14 And & G8III & 2.331 & 28 Dec. 2013 & 4 x 120 & 190 \\
BD +27 2057  & \nodata  & \nodata & G7III & 7.220 & 30 Dec. 2013 & 4 x 900 & 100 \\
\multicolumn{8}{c}{\textbf{Halo}} \\
HD 8724 & 6710 & BD +16 149 & G5 & 5.636  & 31 Dec. 2013 & 4 x 500 & 195 \\
HD 63791 & 38621 & BD +62 959 & G0  & 5.427  &  1 Jan. 2014  & 4 x 900  & 300 \\
\enddata
\tablecomments{Spectral types and $K$ magnitudes from SIMBAD. Sixteen stars have probabilities greater than 90\% of belonging to their predicted population, while seven stars are in the range 0.75 $\leq$ $P$ $<$ 0.9 (HD 29065, thin; HD 37171, thick; HD 49520, thin; BD +27 2057, thick; HD 6555, thin; HD 106760, thin; and HD 211683, thin), and four stars are in the range 0.5 $<$ $P$ $<$ 0.75 (HD 10057, thin, at $P$=0.52; HD 50778, thick, at $P$=0.56; HD 3546, thick, at $P$=0.64; and HD 212074, thin, at $P$=0.72). While HD 10057 is designated as a thin-disk member with $P$ = 0.52, it was identified as a thick-disk giant with a 97\% probability by \citet{Pak13} based on criteria from \citet{Mish04}.}
\end{deluxetable*}

\section{Observations and Data Reduction}\label{obs}

\subsection{Target Selection}
Candidate stars belonging to the Galactic disk were selected for observation from the surveys of \citet{BEM89}, \citet{McW90}, \citet{AB10}, \citet{RAM+1}, \citet{PVK11}, \citet{Wu11}, and \citet{Pak13}. UVW space velocities were computed using the PyAstronomy Galactic space velocity function\footnote{https://pyastronomy.readthedocs.io/en/latest/pyaslDoc/aslDoc /gal\_uvw.html} with the peculiar solar motion correction from \citet{Ding19}. Coordinates, proper motions, distances, and radial velocities were taken from the Gaia Data Release 3 (DR3; \citealt{gaia16, gaia23}). Membership in different kinematic groups was established by requiring that the probability of belonging to one population be greater than 50\%, following the methodology of \citet{JS87} as described by \citet{RAM+1}. The final sample includes nine thin-disk stars, sixteen thick-disk stars, and two halo stars, as shown in Table \ref{table_log}.

\subsection{Observations}
Spectra covering the relatively unblended 2.3359 $\mu$m (1$-$0) R9 feature of HF and the Na I line at 2.3379 $\mu$m were obtained using the Phoenix infrared spectrometer on the 2.1m telescope at the Kitt Peak National Observatory in 2012 November and December, and 2013 December. The spectrometer was configured with a 0."7 slit and the 4308 filter to isolate grating order 32, providing spectral coverage from 2.3285 to 2.3390 $\mu$m.  Observations were obtained at two slit positions for subtraction of a simultaneous sky spectrum and dark current. Hot stars were observed frequently on each night at similar air mass to allow effective telluric-line correction, and stars were observed close to the meridian to mimimize telluric absorption features. 

\subsection{Data Reduction}
\begin{figure}
\epsscale{1.17}
\plotone{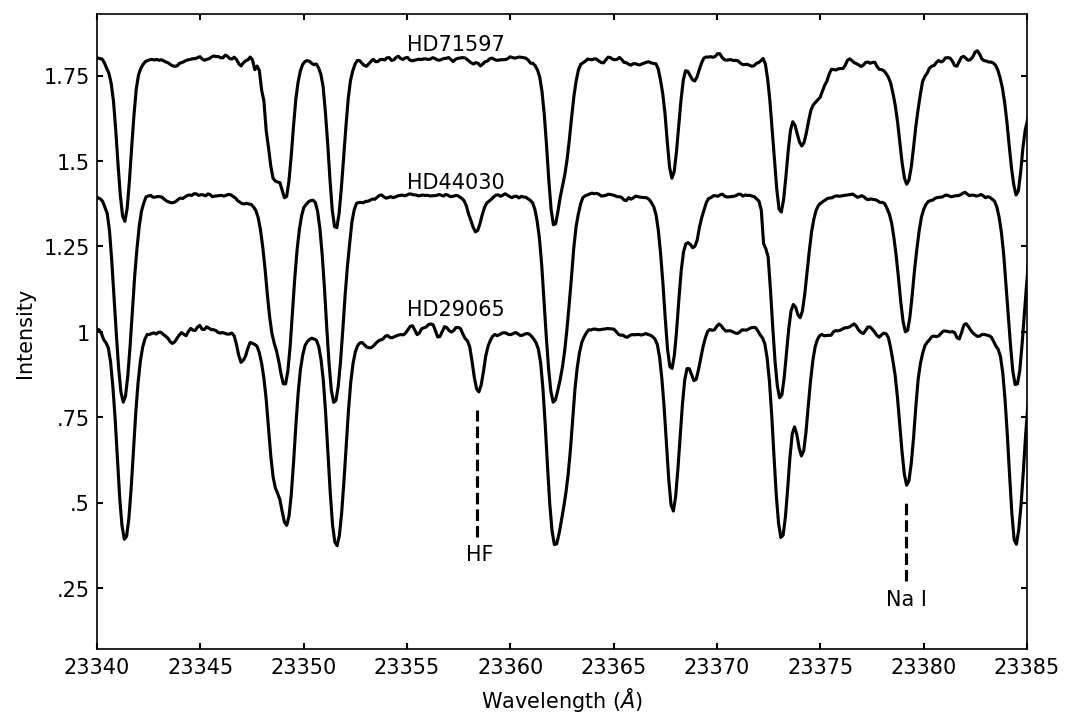}
\caption{Sample spectra showing the features of HF at 23359 \AA \space and Na I at 23379 \AA \space used in this analysis. The top spectrum is HD 71597, the middle spectrum is HD 44030, and the bottom spectrum is HD 29065. The spectra have been shifted vertically for display purposes. Spectra were selected to show the range of HF equivalent width and signal-to-noise ratio used in the analysis.}
\label{sample_spectra}
\end{figure}

Spectra were reduced by subtracting background and night-sky emission using the ABBA observing technique, correcting for variations in pixel sensitivity by dividing by a continuous lamp spectrum, extracting one-dimensional spectra, removing telluric-line contamination by dividing by the spectrum of a hot star, and calibrating the wavelength scale using telluric methane lines in the hot-star spectra. Finally, the spectra were normalized in the continuum and shifted to match the laboratory rest wavelength scale. The observations of each star, including star identifications, spectral types and $K$ magnitudes from SIMBAD, the dates of observation, the total exposure time, and the signal-to-noise (S/N) ratio of the resulting spectra are provided in Table \ref{table_log}. S/N ratios of the sample star spectra range from 100 to 300, while the S/N ratios of telluric-line stars were in excess of 300. Sample spectra of the sample stars are shown in Figure \ref{sample_spectra}.

\section{Analysis}\label{analysis}
\begin{figure*}
\epsscale{1.115}
\plotone{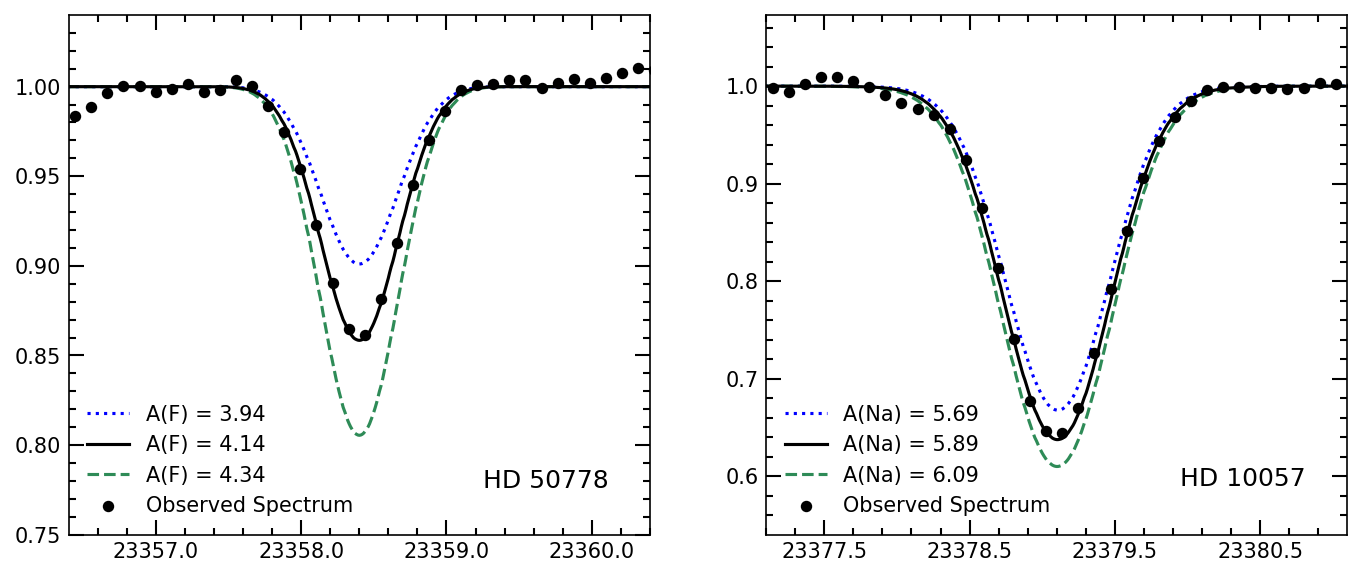}
\caption{The observed and synthetic spectra for the (1$-$0) R9 2.3358 $\mu$m feature of the HF molecule in HD 50778, and the Na I line at 2.3379 $\mu$m in HD 10057. Three syntheses are plotted in each panel, with the best fits being A(F) = 4.14 and A(Na) = 5.89.}
\label{sample_fit}
\end{figure*}

The abundances of fluorine and sodium in the sample stars were determined using spectrum synthesis with the local thermodynamic equilibrium (LTE) spectral analysis code MOOG (V. 2019, \citealt{S73}). Synthetic spectra covering the 2.3359 $\mu$m (1-0) R9 feature of HF and the Na I line at 2.3379 $\mu$m were convolved with Gaussian profiles to fit the final reduced spectra. Model atmosphere parameters (effective temperature, surface gravity, [Fe/H], and microturbulance) were adopted from literature sources identified in Table \ref{table_pars}. Model atmospheres were interpolated from the MARCS grid \citep{G+08} and spherical models were used for our sample of giants. Sample best fit syntheses for the HF and Na I line are shown in Figure \ref{sample_fit}.

\begin{deluxetable*}{lccccccccc}
	\tabletypesize{\scriptsize}
	\tablecaption{Stellar Parameters and Abundances\label{table_pars}}
	\tablewidth{0pt}
	\tablehead{
		\colhead{Star} & \colhead{$T_{eff}$} & \colhead{Log $g$} & \colhead{$\xi$} & \colhead{[Fe/H]} & \colhead{Ref} & \colhead{log $\epsilon$($F$)} & [F/Fe] & \colhead{log $\epsilon$($Na$)} & [Na/Fe] \\
		 & \colhead{(K)} &  & \colhead{(km s$^{-1}$)}
		&  &  &  & & &
	}
	\startdata
        \multicolumn{10}{c}{\textbf{Thin Disk}} \\
        HD 6555  & 4720 & 3.00 & 1.15 & --0.09 &  b  & \nodata & \nodata & 6.40 & 0.25 \\
        HD 10057 & 4130 & 1.70 & 1.35 & --0.30 &  b  & 4.15 & 0.03 & 5.89 & -0.05 \\
        HD 29065 & 3990 & 1.77 & 2.30 & --0.35 &  c  & 4.00 & -0.07 & 5.89 & 0.00 \\
        HD 49520 & 4300 & 2.35 & 2.20 &  0.00 &  c   & 4.20 & -0.22 & 6.24 & 0.00 \\
        HD 106760 & 4510 & 2.70 & 2.10 & --0.29 &  c   & 4.16 & 0.03 & 5.95 & 0.00 \\
        HD 211683 & 4450 & 1.80 & 1.35 & --0.13 &  b   & 4.62 & 0.33 & 6.11 & 0.00 \\
	HD 212074 & 4700 & 2.55 & 1.30 & +0.05 &  b   & \nodata & \nodata & 6.39 & 0.10 \\
	HD 216174 & 4440 & 2.53 & 2.40 & --0.53 &  c   & \nodata & \nodata & 5.71 & 0.00 \\
        HD 233517 & 4475 & 2.25 & 1.85 & --0.37 &  p   & \nodata & \nodata & 5.87 & 0.00 \\
        \multicolumn{10}{c}{\textbf{Thick Disk}} \\
        Arcturus & 4286 & 1.66 & 1.74 & --0.52 & q  & 3.73 & -0.17 & 5.91 & 0.19 \\
        HD 249  & 4850 & 2.96 & 1.17 & --0.15 &  b   & \nodata & \nodata & 6.29 & 0.20 \\
        HD 3546  & 5109 & 2.83 & 1.73 & --0.50 &  i  & \nodata & \nodata & 5.79 & 0.05 \\
        HD 9138  & 4395 & 2.07 & 1.98 & --0.34 &  a  & 4.56 & 0.48 & 6.25 & 0.35 \\
        HD 15596 & 4936 & 3.06 & 1.19 & --0.53 &  i  & \nodata & \nodata & 5.76 & 0.05 \\
        HD 37171 & 4000 & 1.25 & 1.35 & --0.55 &  b  & 3.90 & 0.03 & 5.74 & 0.05 \\
	HD 44030 & 4300 & 1.00 & 3.00 & --0.70 &  d   & 4.23 & 0.51 & 5.79 & 0.25 \\
        HD 50778 & 4084 & 1.70 & 1.50 & --0.22 &  e   & 4.13 & -0.07 & \nodata & \nodata \\
        HD 71597 & 4271 & 1.90 & 1.80 & --0.40 &  f   & $<$3.65 & $<$-0.37 & 5.89 & 0.05 \\
	HD 77729 & 4270 & 2.00 & 3.00 & --0.40 &  d   & 4.20 & 0.18 & \nodata & \nodata \\
	HD 96436 & 4750 & 2.75 & 1.50 & --0.50 &  k   & \nodata & \nodata & 5.94 & 0.20 \\
	HD 103813 & 4816 & 3.11 & 1.01 & --0.36 &  l  & \nodata & \nodata & 6.08 & 0.20 \\
	HD 105740 & 4700 & 2.50 & 1.00 & --0.51 &  m   & \nodata & \nodata & 5.93 & 0.20 \\
        HD 107328 & 4442 & 1.95 & 1.51 & --0.39 &  n   & \nodata & \nodata & 5.85 & 0.00 \\
	HD 121146 & 4400 & 2.25 & 1.90 & --0.06 &  o   & 4.64 & 0.28 & 6.43 & 0.25 \\
        HD 221345 & 4740 & 2.61 & 1.44 & --0.27 &  i   & \nodata & \nodata & 5.97 & 0.00 \\
	BD +27 2057 & 4810 & 2.25 & 1.25 & --0.51 &   g  & \nodata & \nodata & 6.03 & 0.30  \\
        \multicolumn{10}{c}{\textbf{Halo}} \\
	HD 8724  & 4757 & 1.77 & 1.65 & --1.53 &  h  & \nodata & \nodata & 4.41 & -0.30 \\
	HD 63791 & 4725 & 1.70 & 1.60 & --1.68 &  j   & \nodata & \nodata & 4.46 & -0.10 \\
	\enddata
        \tablecomments{We use log $\epsilon$($Fe$)$_\odot\ = 7.50$ from \citet{AGSS}, log $\epsilon$($F$)$_\odot\ = 4.42$ from \citet{Lodders09}, and log $\epsilon$($Na$)$_\odot\ = 6.24$ from \citet{AGSS}.}
	\tablerefs{
		(a) \citet{LH7}; 
		(b) \citet{Pak13}; 
		(c) \citet{McW90}; 
		(d) \citet{GO84}; 
		(e) \citet{dS06}; 
		(f) \citet{BEM89}; 
		(g) \citet{A12};
		(h) \citet{H18};
		(i) \citet{dS15};
		(j) \citet{B00};
		(k) \citet{CS86};
		(l) \citet{DS18};
		(m) \citet{S96};
		(n) \citet{S16};
		(o) \citet{Luck95};
		(p) \citet{Bal+00};
		(q) \citet{ram11}.
	}
\end{deluxetable*}

Following \citet{Ryde20}, we adopt the meteoritic solar fluorine abundance of $A(F)_\odot\ = 4.42$ from \citet{Lodders09} due to the large uncertainties of the solar fluorine values derived from modeling umbral spectra. 

The molecular line data used for analysis of the 23358.329 \AA \space (1$-$0) R9 feature of HF has historically varied. We adopted the excitation potential $\chi = 0.227$ from \citet{decin00}, the HITRAN database \citep{Roth13}, and \citet{Jon14}, log $gf$ = -3.962 from \citet{Jon14}, and dissociation energy of 5.869 from \citet{sauval84}. In early abundance determinations the excitation potential of 0.49 eV was used, which results in a higher fluorine abundance of typically 0.36 (see Section 3.2 in \citealt{NP13}). \citet{Jon14b} confirmed that the partition functions used with MOOG and the \citet{decin00} line data are compatible with their results. Several other log $gf$ values have been determined, such as -3.971 \citep{LUC} and -3.956 \citep{Mai14}, which are similar to the adopted value from \citet{Jon14} and would result in an abundance difference of only $\sim$0.01. Some uncertainty remains in the HF molecule's dissociation energy, which could result in a lower fluorine abundance of $\sim$0.04 dex (see Section 4.1.1 in \citealt{Guer19}). 

Atomic line data for the Na I line at 2.3379 $\mu$m were adopted from \citet{PP}, who used the atomic and molecular line list generator \texttt{linemake} \citep{placco21}, then confirmed by fitting the Na I line in the Infrared Atlas of the Arcturus Spectrum \citep{H+95}.

Multiple CO (2$-$0) and (3$-$1) vibration-rotation lines are found near the HF and Na features. The neighboring C$^{12}$O$^{16}$ lines are included in the synthesis calculation using adopted wavelengths, log($gf$) values, and excitation potentials from \citet{Go94}. Due to the lack of carbon and oxygen abundances for our program stars in the literature, we are unable to measure individual abundances of these elements from only the observed CO lines. The carbon abundance was adjusted to match the observed CO strength, then the abundances of fluorine and sodium were measured in a similar manner.

In one of our program stars (HD 121146), the (2$-$0) R25 line of C$^{12}$O$^{17}$ is present and partially blended with the HF line. This C$^{12}$O$^{17}$ line at 23357.177 \AA \space was included in the synthesis using line data from \texttt{linemake}. \texttt{linemake} also provided line data for the C$^{12}$O$^{17}$ line at 23358.237 \AA \space \citep{mura20}, which was not detected in HD 121146. 

\begin{figure}
\epsscale{1.178}
\plotone{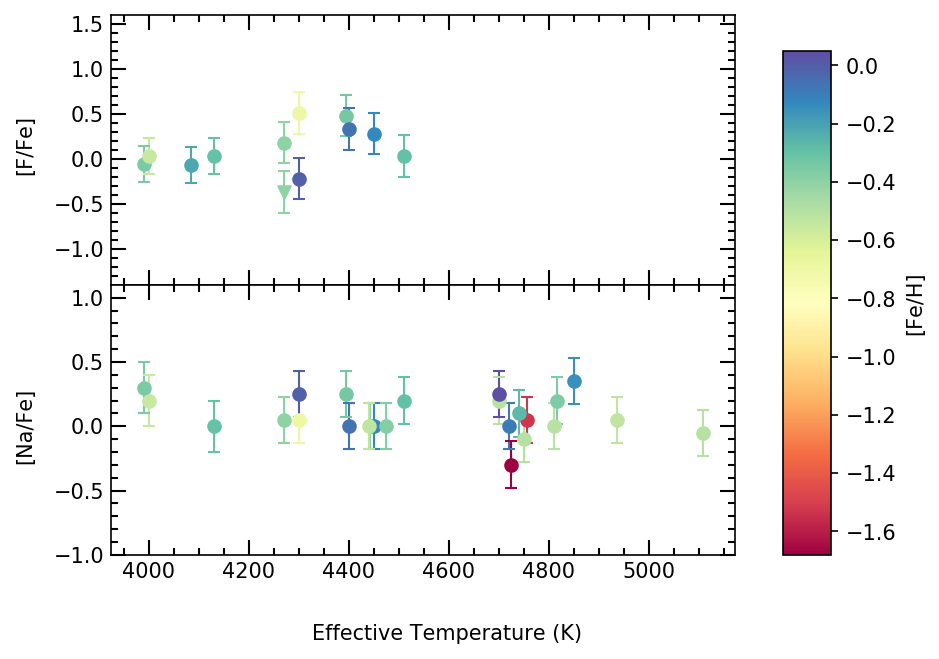}
\caption{Derived abundance vs. effective temperature, with metallicity ([Fe/H]) indicated by color.}
\label{abun_temp}
\end{figure}

Final fluorine and sodium abundance measurements are shown in Table \ref{table_pars}. We report fluorine abundances for five thin-disk and six thick-disk stars, and one thick-disk star upper limit. Stars must be sufficiently cool (less than $\sim$4700 K) to permit the formation of HF in their atmospheres. In our sample stars that are at temperatures above this limit, the HF line was not detected and only measurements of the Na I line were obtained. The [F/Fe] abundance ratios are plotted versus effective temperature in Figure \ref{abun_temp}, with the metallicity shown in the color bar following \citet{Ryde20}. As observed and discussed in \citet{PP} and \citet{Ryde20}, we see a rise in fluorine abundance with temperature that is likely a systematic effect. In the warmer stars, the fluorine feature will only be detected if the fluorine abundance is high, as the HF feature disappears in warmer stars due to molecular dissociation.

Our analytic methods can be confirmed by measuring the 2.3359 $\mu$m (1$-$0) R9 HF feature in the cool giant standard Arcturus from the atlas of \citet{H+95}. We derived a fluorine abundance of A(F) = 3.73 using the atmospheric parameters from \citet{ram11} ($T_{eff}$ = 4286 K, log $g$ = 1.66, [Fe/H] = -0.52, and $\xi$ = 1.74 km s$^{-1}$). This value is comparable with recent Arcturus fluorine measurements by \citet{Guer19} (A(F) = 3.63), \citet{Abia15} (A(F) =  3.78), \citet{Jon14} (A(F) = 3.75), \citet{Jon14b} (A(F) = 3.65 from the 2.3 $\mu$m line and A(F) = 3.73 from the 12.2 $\mu$m line), and \citet{NP13} (A(F) = 3.75). 

We detect rotation in HD 233517, which is a Li-rich, single K giant known to have a high rotation velocity ($v$ sin $i$). By fitting the rotational broadening of the 2.3379 $\mu$m Na I line, we find a $v$ sin $i$ of 17.5 $\pm$ 4 km s$^{-1}$. The rotational line broadening of HD 233517 was previously measured by \citet{Bal+00} and \citet{strass15}, who report values of 17.6 $\pm$ 1.0 km s$^{-1}$ and 17.8 $\pm$ 0.1 km s$^{-1}$, respectively.

No known non-LTE (NLTE) corrections exist for the 2.3359 $\mu$m (1-0) R9 HF feature. LTE should generally be expected for molecular vibrational-rotational transitions in the ground electronic state \citep{Hinkle75}. \citet{li13} explored the deviations from radiative equilibrium for the 2.3359 $\mu$m (1-0) R9 HF line in the low-metallicity range near [M/H] = −2.0. They determined that the effects of three-dimensional (3D) model atmospheres on their derived fluorine abundances were insignificant at low gravity and temperature, but larger corrections ($>0.1$ dex) were expected at higher gravity and temperature. Our sample consists of stars with temperatures less than 4500 K, with the exception of HD 106760, which has an effective temperature of 4510 K, surface gravity of 2.70, and a derived fluorine abundance of [F/Fe] = 0.03. In Figure \ref{abun_temp} we detect no dependence of fluorine abundance on effective temperature, providing some assurance that NLTE and 3D effects are not substantial for stars less than $\sim$4500 K. With the selection effect of the HF line, where the line is detectable at higher effective temperatures if the abundance is high, it is difficult to identify possible NLTE and 3D effects. However, we can examine the derived abundances at the cool end of our sample, which show no obvious increase in abundance that is indicative of such effects.

No NLTE corrections currently exist for the Na I line at 2.3379 $\mu$m. \citet{Cunha15} discuss NLTE calculations for the studied Na I lines in the APOGEE spectral region in the near-IR $H$ band, finding departures from LTE to be minimal ($\leq$ 0.04 dex) for their sample of giants in NGC 6791. Our sample of giants has a similar parameter range as \citet{Cunha15}'s sample, though our metallicities are lower. We observe no temperature trend with derived sodium abundance, as seen in Figure \ref{abun_temp}. 

Uncertainties in our abundance results arise from observational uncertainties and uncertainties in the adopted atmospheric parameters for each star. Observational uncertainties were estimated by altering the smoothing of the synthetic spectrum and the fits themselves until the difference in abundance measurement were distinguishable. To estimate the uncertainties in the adopted stellar atmospheric parameters, the sensitivity of the fluorine abundance to each individual parameter was measured. These uncertainties are summarized in Table \ref{table_uncertainty}. Uncertainties may arise from the telluric correction of the near-IR spectra. Fortunately, H$_2$O only has a few weak lines in the region of interest. The total uncertainty is determined by adding the individual sources in quadrature. Depending on the effective temperature of the star, the total uncertainty for the derived fluorine abundances is less than 0.23 dex across the temperature range from 4000 to 4500 K.

\begin{deluxetable*}{lrrrrrr}
\tabletypesize{\scriptsize}
\tablecaption{Sensitivity to Stellar Parameters\label{table_uncertainty}}
\tablewidth{0pt}
\tablehead{
\colhead{Parameter} & \colhead{log $\epsilon$(F)}  & &  \colhead{log $\epsilon$(Na)} & & & \\
& \colhead{$T_{eff}$= 4000 K} & \colhead{$T_{eff}$= 4500 K}
& \colhead{$T_{eff}$= 4000 K} & \colhead{$T_{eff}$= 4500 K}
}
\startdata
$\Delta$$T_{eff}$ = $\pm$100 K & 0.18 & 0.22 & 0.14 & 0.12\\
$\Delta$Log $g$ = $\pm$0.3 & 0.02 & 0.00 & 0.00 & 0.00 \\
$\Delta$$\xi$ = $\pm$0.5 km s$^{-1}$ & 0.02 & 0.00 & 0.12 & 0.10 \\
$\Delta$[Fe/H] = $\pm$0.1 & 0.05 & 0.03 & 0.06 & 0.07 \\
Observational Uncertainty & 0.05 & 0.05 & 0.05 & 0.05 \\
Total Uncertainty & 0.20 & 0.23 & 0.20 & 0.18 \\
\enddata
\end{deluxetable*}

\section{Comparison to Previous Work}\label{comparison}

To our knowledge, we present the first fluorine measurements for our sample stars. Figure \ref{metallicity} shows the derived abundances versus metallicity, compared to normal giants from the literature. Fluorine abundances from \citet{li13}, \citet{Jon14b}, \citet{PP}, \citet{Guer19}, \citet{Ryde20}, and the data from \citet{Jon17} that \citet{Ryde20} reanalyzed are included. The measurements from these studies have been scaled to align with the molecular data and solar fluorine abundance used in this study. The derived sodium abundances are also presented in Figure \ref{metallicity}, with sodium abundances of red-giant stars in the thin/thick disk and halo from \citet{AB10} for comparison.

Our sample covers a metallicity range from $-$0.70 $<$ [Fe/H] $<$ 0.00, with measurements at low- and approximately-solar metallicity. Recent fluorine abundance studies with measurements in this regime include \citet{Jon14b} ($-$0.62 $<$ [Fe/H] $<$ $-$0.14), \citet{PP} ($-$0.6 $<$ [Fe/H] $<$ +0.3), \citet{Guer19} ($-$1.3 $<$ [Fe/H] $<$ 0.00), and \citet{Ryde20} ($-$1.11 $<$ [Fe/H] $<$ +0.37). For our sample, the average fluorine abundance is [F/Fe] = +0.14 $\pm$ 0.07 (standard error of the mean, SEM) with a standard deviation of 0.23. The \citet{PP} sample of $\sim$40 normal giants has an average abundance of [F/Fe] = +0.07 $\pm$ 0.04 (SEM). Average fluorine abundances in \citet{Guer19} and \citet{Ryde20} are [F/Fe] = $-$0.10 $\pm$ 0.06 (SEM) and [F/Fe] = +0.03 $\pm$ 0.02 (SEM), respectively. By taking an average of the fluorine abundances from the 2.3$\mu$m and 12.2$\mu$m where measurements exist from each line, the average fluorine abundance from \citet{Jon14b} is [F/Fe] = $-$0.09 $\pm$ 0.03 (SEM).

At approximately-solar metallicity, the abundance measurements from all studies have similar [F/Fe] values, though there is an observed scatter (see Figure \ref{metallicity}). The scatter is greater at solar metallicity than at supersolar metallicities. As seen in Table \ref{table_uncertainty}, the dominant source of uncertainty on [F/Fe] ratios is from errors in the effective temperature. The methods of determining temperature among different studies may partially explain the observed scatter and differences in average fluorine abundances across different samples. However, fluorine and sodium have similar overall uncertainties due to atmospheric parameters, but the observed scatter in the fluorine abundances is greater than the scatter in the sodium abundances, as seen in Figure \ref{metallicity}. 

Previously, it has been suggested that fluorine in higher-metallicity stars is dominated by a secondary source (see Section 5.1 in \citealt{Ryde20} for an overview on primary and secondary fluorine production sites and processes). \citet{Ryde20} found a secondary behavior in their sample stars and concluded that AGB stars are the main channel of formation in the metallicity range of $-$0.6 $<$ [Fe/H] $<$ 0. This behavior is demonstrated in the A(F) versus A(Fe I) plot in Figure 4 of \citet{Guer22}, where the slope indicative of secondary yields is shown to fit the observed trend of fluorine at higher metallicities. Our derived fluorine abundances in the metallicity range of $-$0.5 to 0.0 are consistent with the values from the literature, following the observed secondary behavior of disk stars. 

At lower metallicities, a primary source of fluorine dominates \citep{Guer19, Ryde20, Guer22}. \citet{Guer19} proposes primary behavior at [Fe/H] $<$ $-$0.5, and secondary behavior at [Fe/H] $>$ $-$0.5. \citet{Ryde20} suggests at lower metallicities the abundance ratios may decrease more slowly than at higher metallicities, and \citet{Ryde20} and \citet{Guer19} report a near-flat trend of abundance with [Fe/H] at low metallicity. \citet{Guer19} reports a gradient of [F/Fe] for their probable thick-disk/halo stars to have a shallow slope of 0.02 $\pm$ 0.03 dex. We report two fluorine measurements in low-metallicity ([Fe/H] $<$ $-$0.5 dex) stars. Our two measurements have derived fluorine abundances that are at least $\sim$0.3 dex greater than the values derived in low-metallicity stars in \citet{Guer19} and the upper limits from \citet{Ryde20}. Our abundance determination for HD 44030, which has a [Fe/H] = $-$0.70, is comparable to the abundance determinations in the low-metallicity regime for the two metal-poor red giants from \citet{Guer19}, determined to be probable members of the Monoceros overdensity. The other abundance measurement at low metallicity for the star HD 37171 at [Fe/H] = $-$0.55 is comparable to measurements from \citet{PP} and the data from \citet{Jon17} that \citet{Ryde20} reanalyzed at a similar metallicity. 

\section{Fluorine Enrichment in the Thin and Thick Disks}\label{disks}

\begin{figure*}
\epsscale{1.178}
\plotone{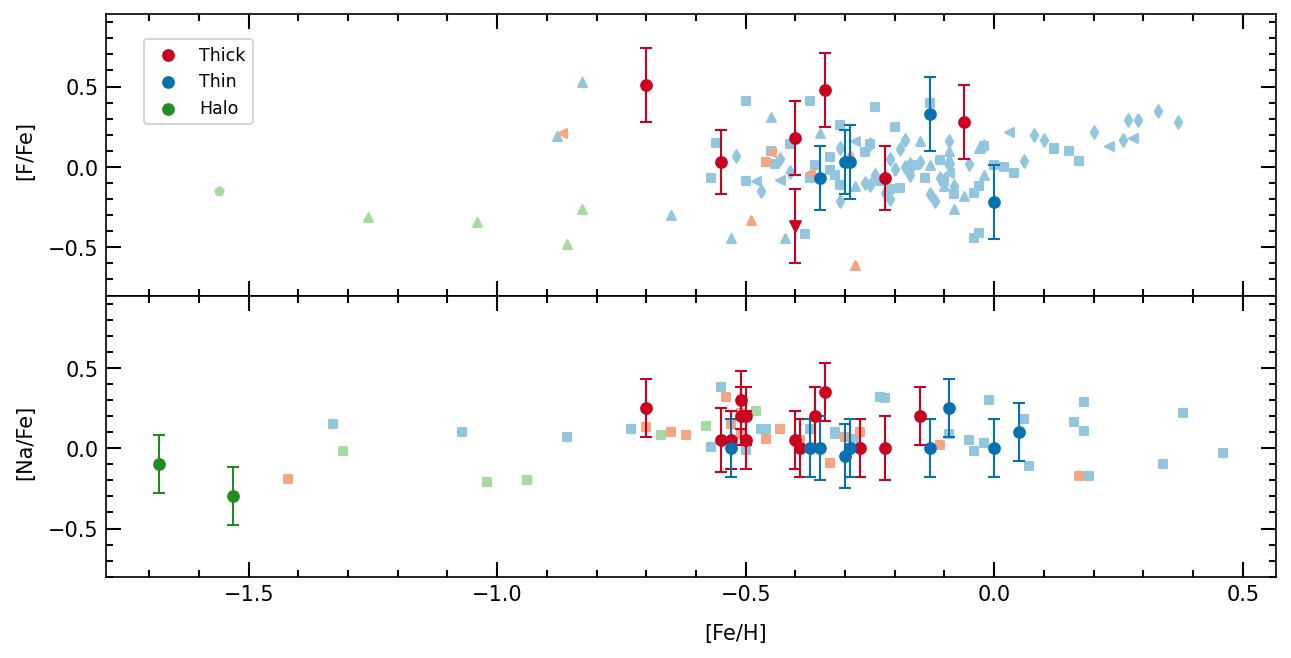}
\caption{[F/Fe] (upper panel) and [Na/Fe] (lower panel) vs. [Fe/H] for thick disk (red circles), thin disk (blue circles), and halo (green circles) stars. Literature measurements for the thin-disk, thick-disk, and halo populations are shown in light blue, orange, and light green, respectively. Abundances for giants are taken from \citet{li13} (pentagons), \citet{Jon14b} (right-pointing arrows), \citet{PP} (squares), \citet{Guer19} (triangles), \citet{Ryde20} (left-pointing arrows), and the \citet{Jon17} data reanalyzed by \citet{Ryde20} (diamonds) for [F/Fe] and from \citet{AB10} (squares) for [Na/Fe].}
\label{metallicity}
\end{figure*}

\begin{figure}
\epsscale{1}
\plotone{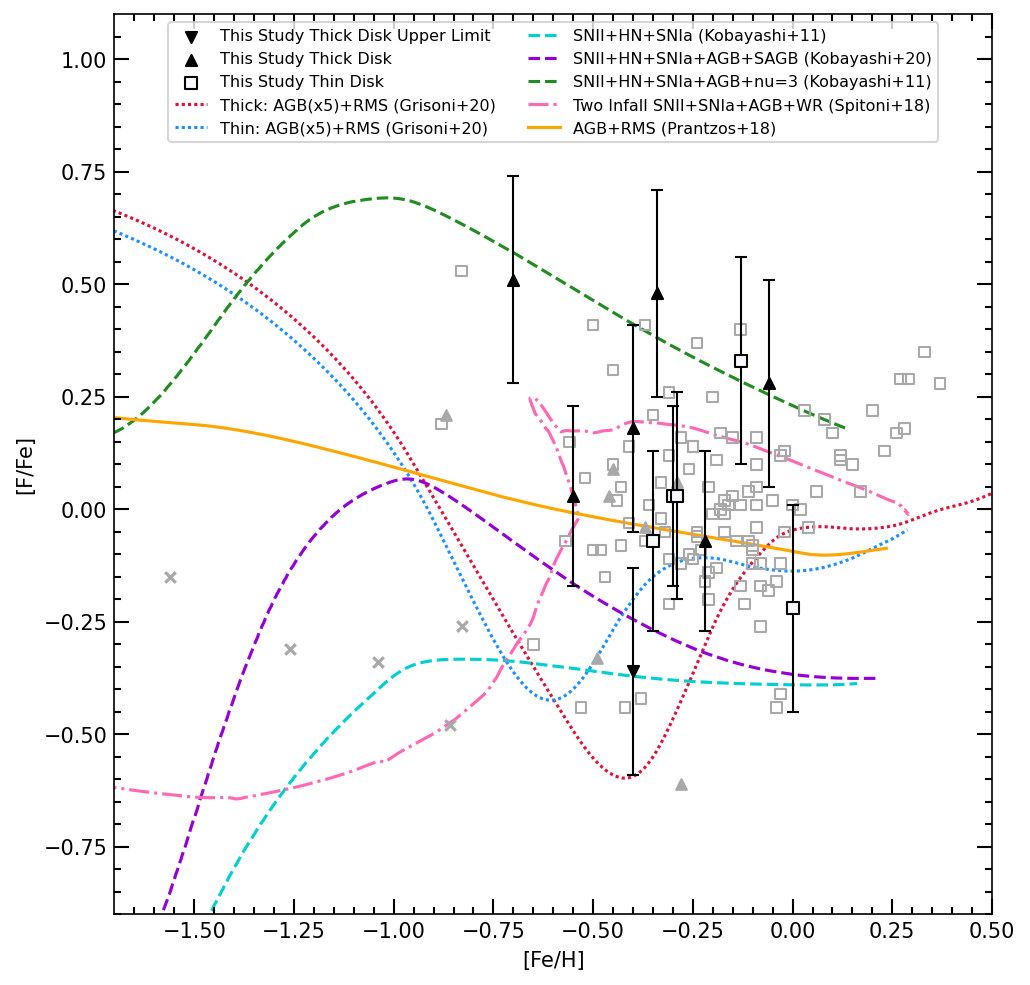}
\caption{[F/Fe] vs. [Fe/H] for our sample of thin- and thick-disk stars with measurements from the literature. Chemical evolution models from the literature are shown for different combinations of production channels. Literature measurements from \citet{li13}, \citet{PP}, \citet{Guer19}, and \citet{Ryde20} are shown in gray. Thin-disk stars are indicated with open squares, thick-disk stars with filled triangles, and halo stars with crosses.}
\label{theoretical}
\end{figure}

Using our classification procedure for kinematic populations detailed in Section 2.1, we assigned the stars from the literature to be members of the thin disk, thick disk, or halo. Gaia DR3 data were utilized when available. In instances where Gaia DR3 data were unavailable, proper motions, parallaxes, and radial velocities came from Gaia DR2 \citep{Gaia18}, \citet{Kar04}, \citet{Fam05}, \citet{Gont06}, \citet{van07}, \citet{Mass08}, \citet{Fam09}, \citet{Soubiran18}, and \citet{Jonsson20}. The final thin-/thick-disk and halo samples from the literature are plotted with our results in Figures \ref{metallicity}, \ref{theoretical}, and \ref{theoretical_oxygen}. Combining our derived abundances with the values from the literature, we find a final thin-disk fluorine abundance of [F/Fe] = +0.04 $\pm$ 0.02 (SEM) with a standard deviation of 0.24, and a thick disk abundance of [F/Fe] = +0.07 $\pm$ 0.08 (SEM) with a standard deviation of 0.29. For thin-disk stars in the metallicity range --0.6 $<$ [Fe/H] $<$ 0.3, \citet{PP} find an average [F/Fe] ratio of +0.07 $\pm$ 0.04 (SEM), which is comparable to our results. Within the range from approximately $-$0.7 $<$ [Fe/H] $<$ 0, the thin- and thick-disk average abundances do not differ by more than 0.03 dex. 

Both our sample and our sample combined with the literature values show a slightly larger dispersion in the thick disk than the thin disk, with standard deviations larger in thick-disk samples. This can be seen in Figures \ref{metallicity} and \ref{theoretical}, and is even more pronounced when considering our upper-limit thick-disk measurement for HD 71597. Scatter in [F/Fe] is higher at lower metallicity, which suggests there could be a variation from star to star in the thick disk. The dispersion could be either real or observational as scatter increases as the HF line gets weaker. The [Na/Fe] in Figure \ref{metallicity} shows a relatively flat trend for both the thin and thick disks with the populations overlapping each other, similar to the distribution and pattern observed by \citet{Li18}.

\section{Chemical Evolution of Fluorine}\label{evolution}

Finally, we compare the observed fluorine abundances to chemical evolution models to constrain possible nucleosynthetic sources of fluorine. Figure \ref{theoretical} shows our derived fluorine abundances and the kinematically assigned literature sample with recent chemical evolution models fully described in \citet{Kob+a,Kob+20}, \citet{Pran18}, \citet{Spit18}, and \citet{gris20} overlain. As in Figure \ref{metallicity}, the literature abundance measurements and theoretical models have been scaled to align with the molecular data and solar fluorine abundance used in this study. 

Some models show a peak in [F/Fe] at $\sim$$-0.1$ which is explained by different star-formation histories. These histories can lead to different abundance patterns, which can be understood in terms of the time-delay model \citep{matt21}. The \citet{gris20} models use a parallel approach (see \citealt{gris17,gris18}), where the disks are formed in parallel, but the thick-disk track is shifted to higher metallicities than the thin disk as a result of its quicker evolution. At solar metallicity, the majority of models show declining [F/Fe] with [Fe/H]. The ``two-infall" model from \citet{Spit18} matches our derived abundances at low metallicity and solar metallicity quite well. Nonetheless, the data show an increase in [F/Fe] at $\sim$+0.25 which is unexplained by the models. The \citet{gris20} models have the closest slope to the observed fluorine abundances from the literature at high metallicity.

\begin{figure}
\epsscale{1}
\plotone{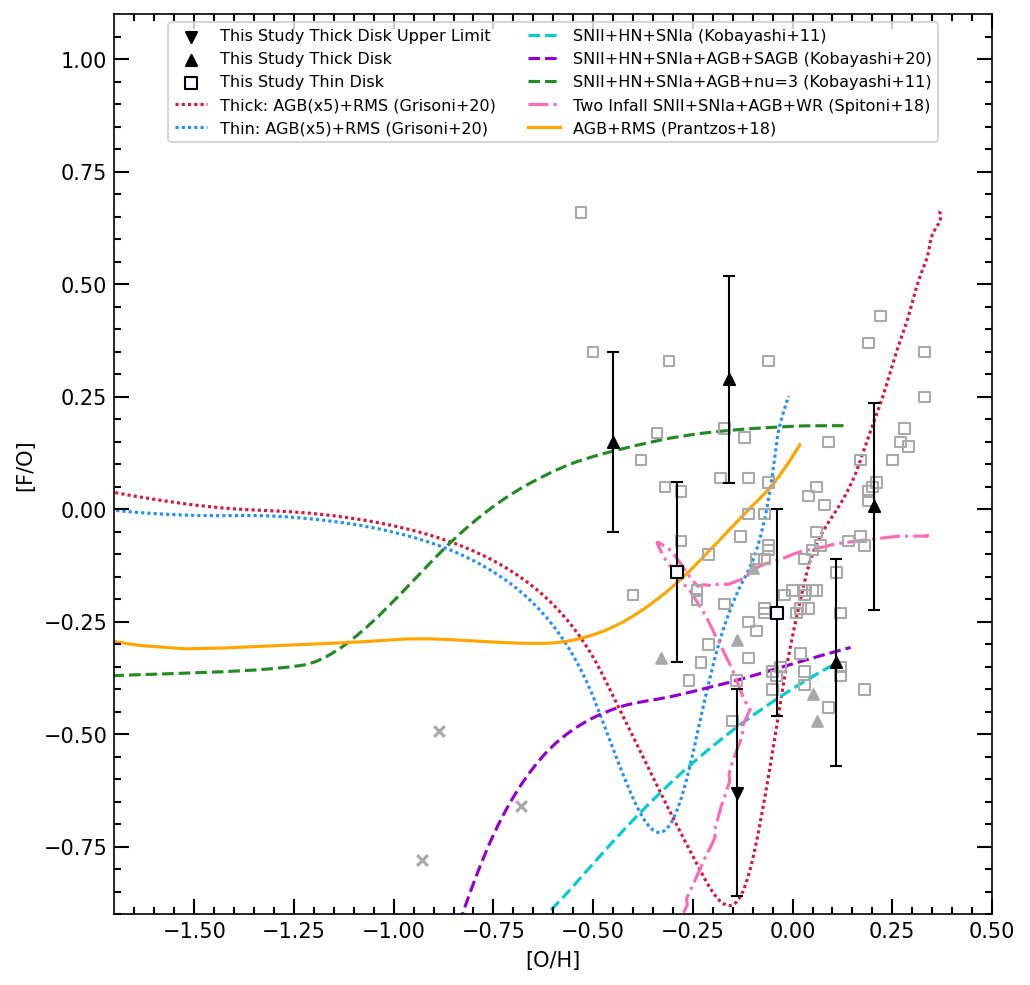}
\caption{[F/O] vs. [O/H] for our sample of thin- and thick-disk stars with measurements from the literature. Symbols are as defined in Figure \ref{theoretical}.}
\label{theoretical_oxygen}
\end{figure}

Thin- and thick-disk stars were distinguished when comparing [F/O] ratios to [O/H] ratios in \citet{gris20}, possibly due to more uncertain yields of Fe. We plot [F/O] vs. [O/H] in Figure \ref{theoretical_oxygen} using oxygen abundances from \citet{Smith90}, \citet{Luck95}, \citet{Mel+08}, \citet{Hinkle14} (Hypatia Catalog), and \citet{Jonsson20} (APOGEE). We do not see differences between populations, though this may be due to higher uncertainties. At lower metallicities, the thick-disk stars have higher [F/O] measurements than the models. 

While sophisticated chemical evolution models of fluorine have been created, the predictions have yet to reproduce all features of the observed abundances. This may in part be due to uncertainties remaining in Galactic chemical evolution models, in addition to the uncertainty in deriving fluorine abundances. \citet{gris20} point out that the largest uncertainty in Galactic chemical evolution modeling is nucleosynthesis prescriptions, leading to proposed corrections to help explain the observed abundance trends. Many prescriptions may be underestimated: \citet{gris20} multiplied the yields of AGB stars by 5, \citet{Pran18} proposed a factor of 2 increase in AGB star yields, and \citet{Spit18} increased the yields of AGB stars, novae, and Wolf$-$Rayet stars. Moreover, there are different nucleosynthesis prescriptions for AGB stars that provide distinctive results when implemented in chemical evolution models. Figure 1 of \citet{gris20} illustrates how different sets of low- and intermediate-mass star yields result in vastly different fluorine predictions on a plot of [F/Fe] versus [Fe/H]. These nucleosynthesis prescriptions affect the model behavior at high metallicity. Instances where the slope of the literature abundances at higher metallicities is not reproduced by a model may be due to the nucleosynthesis prescriptions used for low- and intermediate-mass stars in that model. Further data and investigation into the prescriptions of different nucleosynthetic sources of fluorine are needed to draw definitive conclusions on the origin of fluorine in different stellar populations.

\section{Summary}\label{summary}
Fluorine and sodium abundances have been determined for a sample of giants in the thin disk, thick disk, and halo. We summarize our findings below. 
\begin{itemize}
    \item Our sample covers a metallicity range from $-$0.70 $<$ [Fe/H] $<$ 0.00. The average fluorine abundance for our sample is [F/Fe] = +0.14 $\pm$ 0.07 (SEM) with a standard deviation of 0.23. The derived fluorine abundances are consistent with other studies of fluorine in these populations.
    \item We report two robust fluorine measurements for stars with [Fe/H] $<$ 0.5 dex: [F/Fe] = +0.51 (HD 44030) and [F/Fe] = +0.03 (HD 37171).
    \item We identify two stars with high [F/Fe]: HD 9138 with [F/Fe] = +0.48 and HD 44030 with [F/Fe] = +0.51.
    \item The thin- and thick-disk average [F/Fe] for our sample of stars combined with the literature differ by 0.03 dex.
    \item There is a larger dispersion in fluorine abundances than sodium abundances despite both species having similar overall uncertainties due to atmospheric parameters, suggesting this dispersion is real and not observational. The dispersion is slightly larger in the thick disk than the thin. 
    \item Finally, we compare our observations with Galactic chemical evolution models including different nucleosynthesis prescriptions for fluorine. Still, it remains difficult to draw firm conclusions and further data might be needed in this context.
\end{itemize}

\section{Acknowledgements }\label{acknowledgements}
We thank the referee for helpful comments that improved this paper. We are grateful to the Kitt Peak National Observatory and particularly to Karen Butler, Ken Hinkle, Dick Joyce, and Christian Soto for their assistance during the observing runs. We also thank Rashad Givens for his contributions in the early phases of the analysis of the data. C.A.P. acknowledges the generosity of the Kirkwood Research Fund at Indiana University. V.G. acknowledges financial support from INAF under the program “Giovani Astrofisiche ed Astrofisici di Eccellenza - IAF: Astrophysics Fellowships in Italy" (Project: GalacticA, “Galactic Archaeology: reconstructing the history of the Galaxy"). Z.G.M is partially supported by a NASA ROSES-2020 Exoplanet Research Program Grant (20-XRP20 2-0125). This research has made use of the NASA Astrophysics Data System Bibliographic Services, the HITRAN database operated by the Center for Astrophysics, and the WEBDA database, operated at the Department of Theoretical Physics and Astrophysics of the Masaryk University. Additionally, this research has made use of the SIMBAD database, operated at CDS, Strasbourg, France. The research shown here acknowledges use of the Hypatia Catalog Database, an online compilation of stellar abundance data as described in Hinkel et al. (2014), which was supported by NASA's Nexus for Exoplanet System Science (NExSS) research coordination network and the Vanderbilt Initiative in Data-Intensive Astrophysics (VIDA). Finally, this work has made use of data from the European Space Agency (ESA) mission
{\it Gaia} (\url{https://www.cosmos.esa.int/gaia}), processed by the {\it Gaia} Data Processing and Analysis Consortium (DPAC,
\url{https://www.cosmos.esa.int/web/gaia/dpac/consortium}). Funding for the DPAC has been provided by national institutions, in particular the institutions participating in the {\it Gaia} Multilateral Agreement.

%

\vspace{5mm}
\facilities{KPNO:2.1m (Phoenix)}


\software{IRAF \citep{tody86, tody93}, MOOG \citep[][V. 2019]{S73}, matplotlib \citep{hunter07}, numpy \citep{vanderwalt11}}

 \clearpage
 
\end{document}